\documentclass{emulateapj}
\usepackage{graphicx}
\begin{document} 

\title{Globular Cluster Scale Sizes in Giant Galaxies:  The Case of M87 and the Role of Orbital Anisotropy and Tidal Filling}
\author{Jeremy J. Webb, Alison Sills, William E. Harris}
\affil{Department of Physics and Astronomy, McMaster University, Hamilton ON L8S 4M1, Canada}
\email{webbjj@mcmaster.ca}

\begin{abstract}

We present new Hubble Space Telescope imaging of the outer regions of M87 in order to study its globular cluster (GC) population out to large galactocentric distances. We discuss particularly the relationship between GC effective radii $r_h$ and projected galactocentric distance $R_{gc}$. The observations suggest a shallow trend $r_h \propto R_{gc}^{0.14}$ out to $R_{gc} \sim 100$ kpc, in agreement with studies of other giant elliptical galaxies. To theoretically reproduce this relationship we simulate GC populations with various distributions of orbits. For an isotropic distribution of cluster orbits we find a steeper trend of $r_h \propto R_{gc}^{0.4}$. Instead we suggest that (a) if the cluster system has an orbital anisotropy profile, where orbits become preferentially radial with increasing galactocentric distance, \textit{and} (b) if clusters become more tidally under-filling with galactocentric distance, the observed relationship can be recovered. We also apply this approach to the red and blue GC populations separately and predict that red clusters are preferentially under-filling at large $R_{gc}$ and have a more isotropic distribution of orbits than blue clusters.
\end{abstract}


\section{Introduction \label{Introduction}}

Many globular cluster (GC) properties, even simple ones like scale size, lack fundamental explanations. It is typically assumed that the gravitational field of the host galaxy is responsible for limiting cluster size \citep[e.g.][]{vonhoerner57, king62, innanen83, jordan05, binney08, bertin08}. The tidal field imposes a \textit{tidal radius} $r_t$, also known as the Jacobi radius $r_J$, of the GC, beyond which a star feels a stronger acceleration towards the host galaxy than toward the cluster and can escape. It is often assumed that the observationally determined \textit{limiting radius} $r_L$, which marks the point where cluster density drops to zero \citep{binney08}, represents $r_t$. But comparisons of the observational relationship between cluster size and galactocentric distance to theory are beginning to suggest otherwise.

First-order tidal theory suggests that the $r_t$ of a GC on a circular orbit is related to its galactocentric distance \citep{vonhoerner57} via:

\begin{equation}\label{rtHoerner}
r_t=r_{gc}(\frac{M}{2M_g})^{1/3}
\end{equation}

\noindent where $r_{gc}$ is the three dimensional galactocentric distance of the cluster, M is the the cluster's mass, and $M_g$ is the mass of the galaxy.  Assuming the mean cluster mass is independent of galactocentric distance and the host galaxy potential can be approximated by an isothermal sphere ($M_g(r_{gc}) \propto r_{gc}$), we expect $r_t \propto r_{gc}^{\frac{2}{3}}$. Furthermore, if central concentration \textit{c} is also independent of $r_{gc}$, the mean effective (or half-mass) radius $r_h$ will also be related to galactocentric distance by the same scaling.

Suppose we assume more generally that $r_h \propto R_{gc}^\alpha$, where now $R_{gc}$ is the two-dimensional (projected) galactocentric distance. If $r_h \propto r_{gc}^{\frac{2}{3}}$, then the effects of projection from 3D to 2D would make $\alpha \sim 0.4 - 0.5$ for normal radial distributions. However, observations of GCs in different galaxies do not match this simple theoretical prediction. The Milky Way cluster population comes the closest with $\alpha=0.46 \pm 0.05$ (data from Harris 1996 (2010 Edition)). The discrepancy in $\alpha$ can perhaps be attributed to the Milky Way's non-spherical potential, and to the fact that GCs do not have circular orbits \citep[e.g.][]{dinescu99, dinescu07, dinescu13}. 

Issues due to a non-spherical potential can be minimized by focussing on giant elliptical galaxies. However recent measurements of $\alpha$ in giant elliptical galaxies present an even larger discrepancy between theoretical and observational values. \citet{spitler06} found $\alpha=0.19 \pm 0.03$ for NGC 4594, as did \citet{harris10}.  \citet{gomez07} found relationships for the metal poor and metal rich GC populations of NGC 5128 separately, with $\alpha = 0.05 \pm 0.05$ for the metal poor clusters and $\alpha = 0.26 \pm 0.06$ for the metal rich clusters. \citet{harris09a} found an extremely flat relationship $\alpha = 0.11$ for a sample of six massive gE galaxies. However, \citet{blom12} found that NGC 4365 has a rather steep value of $\alpha$ equal to $0.49 \pm 0.04$ compared to other giant ellipticals, in closer agreement with simple theory. Such a high value of $\alpha$, along with the identification of three distinct GC sub-populations, may indicate NGC 4365 underwent unique stages of formation and evolution compared to the other galaxies mentioned above.

In summary, measurements of $\alpha$ in most galaxies so far yield an observed relationship between $r_h$ and $R_{gc}$ much shallower than predicted. Attempts to explain this disagreement have been inconclusive. \citet{madrid12} used $N$-body simulations to illustrate the relationship between $r_h$ and $R_{gc}$ is better represented by $r_h \propto tanh(R_{gc})$ for identical model clusters on a range of circular orbits in a Milky-Way like potential. They found that $r_h$ increases steadily with galactocentric distance out to 40 kpc, beyond which $r_h$ stays relatively constant as the effect of tides becomes less and less important. However, their model clusters had larger effective radii than clusters seen in the outer regions of giant E galaxies. Application of this work to the potentials of giant E galaxies and including a larger range of non-circular orbits is promising. 

It may instead be the case that outer halo clusters originally formed tidally under-filling, and are still in the process of expanding \citep[e.g.][]{gieles10, webb13}. \citet{strader12} also found that clusters in NGC 4649 showed no relationship between $r_h$ and $R_{gc}$ beyond 15 kpc, indicating they are not tidally truncated. For tidally under-filling clusters, $r_L$ would be distinctly less than the theoretically allowed $r_t$, and tidal theory would then over-estimate their size. It may also be possible that the current location of outer GCs is not indicative of their location when they formed; they may represent a captured population from smaller satellite galaxies. Therefore it would be the cluster's orbit in the potential of the satellite galaxy that first imposed cluster size, making any predictions with the potential of the current host galaxy inapplicable. The concept of GC populations consisting of one or more captured sub-populations has also been used to explain their observed bi-modal or even tri-modal distribution in colour, typically attributed to differences in cluster metallicity \citep[e.g.][]{zepf93, larsen01, brodie06, peng06, harris09a, blom12}.

In a previous paper \citep{webb12} we measured the size distribution of GCs with $R_{gc} \le 10$ kpc in M87, and found a very shallow trend $\alpha = 0.08 \pm 0.02$. We explained the distribution by introducing an anisotropy gradient in the cluster orbits. If GC orbits become more and more radial with galactocentric distance, the mean cluster size will drop below the theoretical prediction as clusters will be subject to increased tidal stripping \citep{webb13} and will thus flatten the relationship between $r_h$ and $R_{gc}$. Unfortunately our work was limited by the range in $R_{gc}$ of our observations. In this paper we present new Hubble Space Telescope (HST) observations of the outer regions of M87, extending cluster size measurements beyond 100 kpc. M87 contains the largest easily accessibly GC population, making it easiest to trace out to large $R_{gc}$. The larger range in $R_{gc}$ allows for a much stronger test of how orbital anisotropy effects the size distribution of GCs.

In Section 2 we introduce our new observations and determine the effective radii of each cluster in order to extend the observed trend between cluster size and galactocentric distance. In Section 3 we discuss the model originally used in \citet{webb12} for simulating a theoretical M87 cluster population, and discuss in detail the improvements we have made. The model makes use of the known mass distribution of M87 \citep{mclaughlin99} and various cluster population parameters (set to match the observations) to establish a theoretical relationship between cluster size and $R_{gc}$. In Section 4 we discuss the comparison between between theory and observations, as well as planned future work.

\section{Observations \label{stwo}}

\begin{figure} 
\centering
\includegraphics[width=\columnwidth]{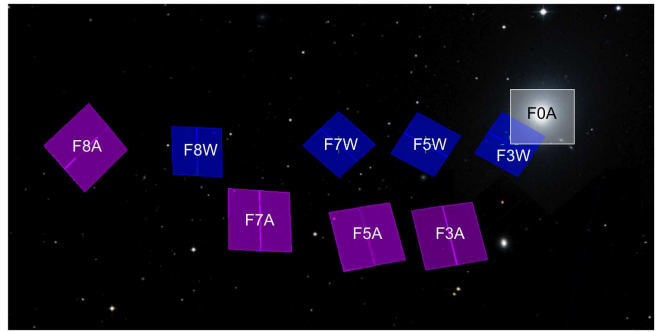}
\caption{Fields of view for new HST images relative to the center of M87. WFC3 images (F3W, F5W, F7W, F8W) are marked in blue and ACS images (F3A, F5A, F7A, F8A) marked in purple. Field of view of archive ACS images (F0A) is marked in white.}
  \label{fig:hstfov}
\end{figure}

We use a combination of archived and new HST images to study the GC population of M87. The new HST images presented in this study are from program GO-12532 (PI Harris), and consist of both Wide Field Camera 3 (WFC3) and Advanced Camera for Surveys (ACS) images of the outer regions ($R_{gc} > 10$ kpc) of M87, extending out to nearly 110 kpc.  For each field, three WFC3 exposures totalling approximately 2600 seconds and three ACS exposures totalling approximately 2300 seconds were taken simultaneously with the F814W filter. The following orbit repeated the same observations with the F475W filter. The process was repeated for three additional ACS/WFC3 pairs for a total of 8 fields of view over 8 orbits. The final co-added composite exposures in each filter were constructed through use of the STSDAS/MULTIDRIZZLE routine within IRAF. The details of of each M87 image are summarized in Table \ref{table:hst} and the locations of our fields are illustrated in Figure \ref{fig:hstfov}. 

\begin{table}
  \caption{HST Image Information}
  \label{table:hst}
  \begin{center}
    \begin{tabular}{lccccc}
      \hline\hline
      {Field} & {RA} & {DEC} & {Camera} & {Filter} & {Exposure Time} \\
       {} & {(J2000)} & {J2000)} & {} & {} & {(seconds)} \\

      \hline

F3WI & 12 30 56.4865 & +12 21 48.20 & WFC3 & F814W & 2589 \\
F3WB & 12 30 56.4865 & +12 21 48.20 & WFC3 & F475W & 2729 \\
F3AI & 12 31 03.691 & +12 27 29.47 & ACS & F814W & 2282 \\
F3AB & 12 31 03.691 & +12 27 29.47 & ACS & F475W & 2351 \\

F5WI & 12 31 15.360 & +12 21 48.30 & WFC3 & F814W & 2589 \\
F5WB & 12 31 15.360 & +12 21 48.30 & WFC3 & F475W & 2729 \\
F5AI & 12 31 23.374 & +12 27 25.58 & ACS & F814W & 2282 \\
F5AB & 12 31 23.374 & +12 27 25.58 & ACS & F475W & 2351 \\

F7WI & 12 31 34.849 & +12 21 48.20 & WFC3 & F814W & 2589 \\
F7WB & 12 31 34.849 & +12 21 48.20 & WFC3 & F475W & 2729 \\
F7AI & 12 31 50.450 & +12 17 13.94 & ACS & F814W & 2282 \\
F7AB & 12 31 50.450 & +12 17 13.94 & ACS & F475W & 2351 \\

F8WI & 12 32 06.642 & +12 21 25.08 & WFC3 & F814W & 2589 \\
F8WB & 12 32 06.642 & +12 21 25.08 & WFC3 & F475W & 2729 \\
F8AI & 12 32 15.130	 & +12 15 50.32 & ACS & F814W & 2282 \\
F8AB & 12 32 15.130 & +12 15 50.32 & ACS & F475W & 2351 \\

      \hline\hline
    \end{tabular}
  \end{center}
\end{table}

We combined our new HST data with archived HST ACS/WFC images of the central regions of M87 in the F814W (I) and F606W (V) filters (also illustrated in Figure \ref{fig:hstfov}), from program GO-10543 (PI Baltz). A detailed description of the co-added composite exposures in each filter can be found in \citet{bird10}. The GCs in this central field have been studied in detail by  \citet{madrid09}, \citet{peng09}, \citet{waters09} and \citet{webb12}. To follow the nomenclature established in Table \ref{table:hst}, these images will be referred to as F0AI and F0AV.

For consistency, our search for GC candidates was performed with the method described in \citet{webb12}. All images were searched for GC candidates with thresholds set such that individual halo stars are rejected while the faintest of GCs are still included. Finally, only objects that were found in both the F814W and F475W filters were accepted, resulting in an initial candidate list of $3287$ objects. 

\subsection{Globular Cluster Effective Radii}

Before we can make any measurements of the structural parameters of our GC candidates, a point spread function (PSF) must first be modelled for each image, which we do empirically. The process is described in detail by \citet{madrid09}. For a given image, stars were identified with SExtractor \citep{bertin96} by approximately measuring the full width half maximum (FWHM) of all objects that are brighter than the background by a factor of 5 times the standard deviation of the background. Star-like objects with FWHMs of approximately 2.5 pixels for the WFC3 images and 2.0 pixels for the ACS images are easily identifiable that correspond to the expected $0\farcs 01$ FWHM of stars. Stars were inspected for faint companions, bad pixels, or other anomalies before use of the standard DAOPHOT routines to build the PSF.

In \cite{webb12}, the surface brightness distribution of each cluster was fit with PSF-convolved \cite{king62} models via the code GRIDFIT  \citep[e.g.][]{barmby07, mclaughlin08, harris10}. Unfortunately, attempts to use GRIDFIT with the new HST dataset resulted in poor fits due to the lower resolution. Therefore we opted to measure the $r_h$ of each cluster candidate with the software ISHAPE \citep{larsen99} which has been successfully used many times on images with similar resolution  \citep[e.g.][]{madrid09}.

\begin{figure} 
\centering
\includegraphics[width=\columnwidth]{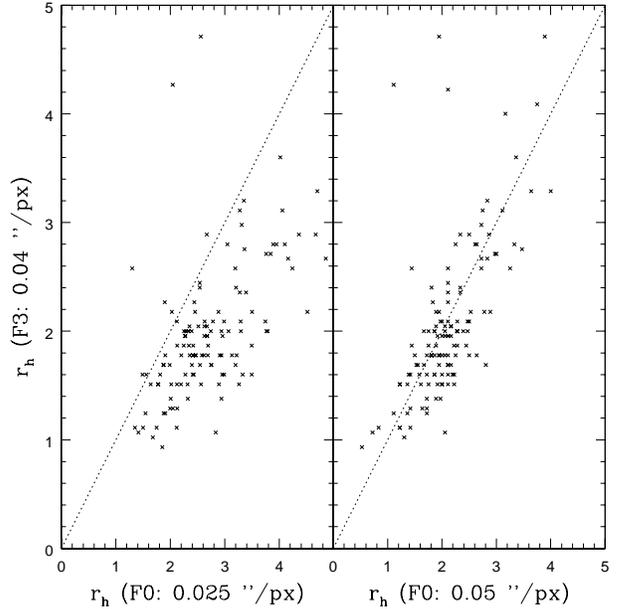}
\caption{$r_h$ of F3W GCs vs. the $r_h$ of overlapping GCs in the high resolution ($0\farcs 025~px^{-1}$) F0A images (left) and low resolution $0\farcs 05~px^{-1}$ F0A images (right). The dotted lines represent a 1:1 correlation.}
  \label{fig:rh_compare}
\end{figure}

For consistency purposes, we also re-measured the GCs in the central field F0A with ISHAPE. We measured these clusters through both $0\farcs 025~px^{-1}$ and $0\farcs 05~px^{-1}$ versions of the F0A combined images. Then, since a portion of the F3W image overlaps with the F0A image, we explore the influence of measuring cluster sizes on images with different detectors by plotting the $r_h$ of clusters found in both images in the left panel of Figure \ref{fig:rh_compare}. From Figure \ref{fig:rh_compare} (left panel), images with lower resolution appear to result in underestimating cluster sizes by a mean value of 0.7 pc, or 0.2 pixels in the lower resolution image. 

To determine whether the discrepancy of 0.7 pc can be attributed to differences in resolution, we compare the GCs in F3 in the right panel of Figure \ref{fig:rh_compare} with the same objects in F0A but now at $0\farcs 05~px^{-1}$. When measured at similar resolutions, the overlapping GCs in each field have comparable effective radii, with the scatter centered around a $1:1$ correlation. The scatter is expected due to the images having significantly different signal-to-noise ratios (F0AV and F0AI images have much longer exposure times equalling 24,500 and 73,800 seconds). Therefore, the mean difference of 0.7 pc in Figure \ref{fig:rh_compare} (left) can be attributed to differences in both resolution and signal-to-noise between the F0A and F3W images. To remain consistent with works of \citet{madrid09}, \citet{peng09}, \citet{waters09} and \citet{webb12} regarding F0A, cluster sizes measured with our new HST dataset in fields F3-F8 are increased by 0.7 pc.

Objects were then removed from the candidate list that were poorly fit by ISHAPE ($\chi^2$ values greater than 10) or that had large differences between the measured $r_h$ in the F814W and F475W bands. For the ACS images, true magnitudes were determined through aperture photometry extrapolated to large radius \citep{sirianni05}. The transformations of \citet{saha11} were then used to convert magnitudes to the standard B and I. However for the WFC3 images, only the filter-based magnitudes could be measured (F475W, F814W) since no well calibrated transformation to (B,I) is available at present. 

\begin{figure} 
\centering
\includegraphics[width=\columnwidth]{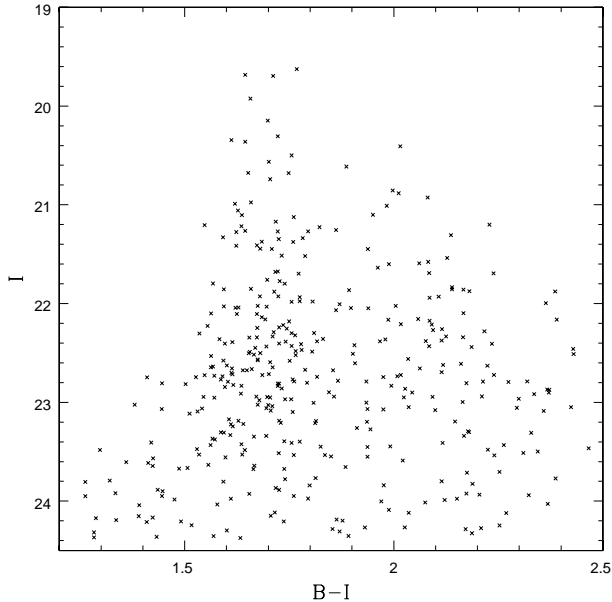}
\caption{CMD of the GC candidates in ACS images of the outer regions M87.}
  \label{fig:HRdiagrama}
\end{figure}

\begin{figure} 
\centering
\includegraphics[width=\columnwidth]{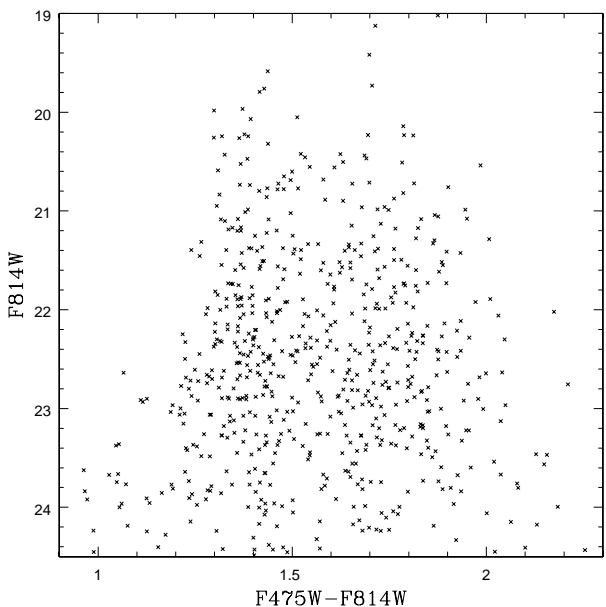}
\caption{CMD of the GC candidates in WFC3 images of the outer regions M87.}
  \label{fig:HRdiagramw}
\end{figure}

The candidate list was trimmed further by cutting objects that were either extremely blue, extremely red, or extremely faint and could be visually identified as non-GCs. Colour-magnitude diagrams (CMDs) of the final 1047 candidates are shown in Figures \ref{fig:HRdiagrama} and \ref{fig:HRdiagramw}. In both CMDs, the blue (metal-poor) and red (metal-rich) sequences are clearly visible. ACS objects with B-I $<$ 1.8 and WFC3 objects with F475W-F814W $<$ 1.5 were declared blue, with the remaining clusters declared red. The size, goodness of fit, colour, and magnitude cuts described above ensure none of the objects in Figures \ref{fig:HRdiagrama} and \ref{fig:HRdiagramw} are either foreground stars or background galaxies.

\begin{figure} 
\centering
\includegraphics[width=\columnwidth]{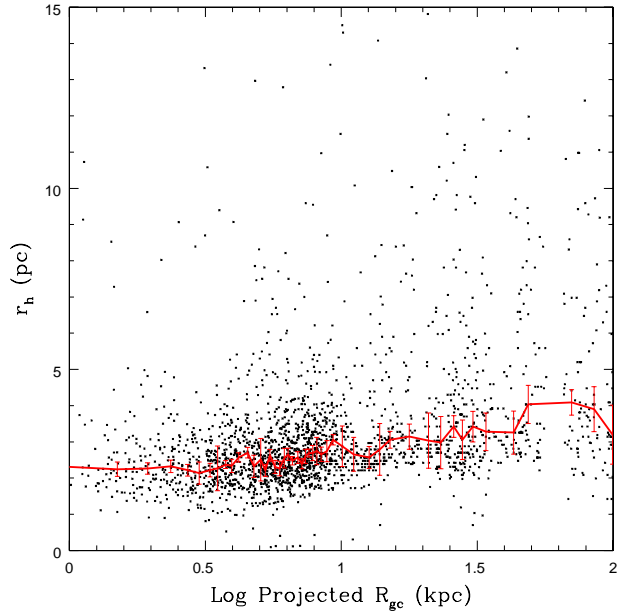}
\caption{$r_h$ vs. log $R_{gc}$ for observed GCs. The solid red line indicates the median $r_h$ calculated with radial bins containing 50 GCs each. Error bars represent the standard error $\sigma/\sqrt(n)$ as given by \citet{harris10}.}
  \label{fig:rh_obs}
\end{figure}

The F814W and F475W images of each field were then co-added to boost the signal to noise ratio, and ISHAPE was again used to measure the $r_h$ of each of the final candidates. The $r_h$ from these combined images as a function of $R_{gc}$ for each candidate is illustrated in Figure \ref{fig:rh_obs}. The median $r_h$ is plotted in red, calculated with radial bins containing 50 GCs each. Finding the slope of the median line in log-log space allowed for the determination of $\alpha$ to be $0.14 \pm 0.01$, similar to the values found in other giant E galaxies discussed in Section 1.

While the relationship between the median $r_h$ and $R_{gc}$ is shallow, it is important to note that the scatter about the median increases with $R_{gc}$. The outer halo of M87 consists of extended ($r_h > 5$ pc) GCs at large $R_{gc}$ and that have been projected to smaller $R_{gc}$. The extended clusters are more in line with what is expected from simple tidal theory, which indicates that the outer halo may comprise a mixture of dynamical histories.

\section{Simulation \label{sthree}}

The observational results shown in Figure \ref{fig:rh_obs} are next compared to models of GCs moving in the tidal field of M87, to constrain our understanding of their scale sizes. The simulation we use is described in detail in \citet{webb12}, but we now extend it further.

\subsection{Initial Conditions}

We first set up a model cluster population with the same observational characteristics (radial profile, velocity dispersion, mass distribution, central concentration distribution) as the observed population of M87 clusters, and then use tidal theory to establish a theoretical relationship between cluster size and $R_{gc}$ that can be compared to Figure \ref{fig:rh_obs}. The simulation allows for separate red and blue populations to be modelled.

Each simulated GC is given a position in the halo ($r$, $\theta$, $\phi$), velocity ($v_r$, $v_\theta$, $v_\phi$), mass (M), and central concentration ($c=\frac{r_t}{r_c}$). The spatial distribution of the red and blue cluster subpopulations is taken from \citet{harris09b}, and we assume the angular distribution to be spherically symmetric. The luminosity function of the F0A GCs, a Gaussian with a mean visual magnitude of -7.3 and a standard deviation of 1.3 \citep{webb12}, is used to establish the mass distribution of GCs with $(\frac{M}{L})_V = 2$ (e.g. \citet{mclaughlin05}). We adopt $(m-M)_0 = 30.95$ for M87 \citep{pierce94, tonry01}. The central concentration of each simulated GC was drawn from the observed distribution of Milky Way clusters from Harris 1996 (2010 Edition), a Gaussian with a mean of $c = 1.5$ and standard deviation of 0.4.

The observed line of sight velocity dispersion ($\sigma$) \citep{cote01} is initially assumed to be identical for each spherical coordinate (R, $\theta$, $\phi$), such that $\sigma_R = \sigma_\theta = \sigma_\phi$. This assumption results in an isotropic distribution of orbits and the anisotropy parameter ($\beta$) equal to zero (Equation \ref{beta}) \citep{binney08},

\begin{equation}\label{beta}
\beta =1-\frac{\sigma_\theta^2+\sigma_\phi^2}{2 \sigma_R^2}
\end{equation}

 In our simulation $\beta$ is kept as a free parameter, and can also change with galactocentric distance. All distribution parameters are summarized in Table \ref{table:gcsim}.

\begin{table}
  \caption{Simulated Globular Cluster Population Input Parameters}
  \label{table:gcsim}
  \begin{center}
    \begin{tabular}{lcc}
      \hline\hline
      {Parameter} & {Value} \\
      \hline

Radial Distribution & Hubble Profile \\
Blue Population & \\
$\sigma_0$ & 66 arcmin$^{-2}$ \\
$R_0$ & 2.0' \\
a & 1.8 \\
Red Population & \\
$\sigma_0$ & 150 arcmin$^{-2}$ \\
$R_0$ & 1.2' \\
a & 2.1 \\
Angular Distribution & Spherically Symmetric \\
Mass-To-Light Ratio & $(M/L)_V$ = 2 \\
Mass Distribution & Gaussian \\
$\langle log(M/M_0) \rangle$ & 5.5 \\
$\sigma_{log(M/M_0)}$ & 0.52 \\
Velocity Dispersion & Gaussian \\
$\langle v \rangle$ & -19 km/s \\
$\sigma_v$ & 401 km/s \\
$\beta$ & 0 \\
Central Concentration & Gaussian \\
$\langle c \rangle$ & 1.5 \\
$\sigma_c$ & 0.4 \\  
      \hline\hline
    \end{tabular}
  \end{center}
\end{table}

While the initial setup of our model population is the same as in \citet{webb12}, improvements have since been made towards making the model cluster population more realistic and representative of the observations we are trying to duplicate. 

\subsection{Calculating Tidal and Effective Radii}

After a cluster has been assigned a position, velocity, mass, and central concentration the orbit of the cluster is then solved. Now we have all the ingredients necessary to calculate $r_t$ and $r_h$,which is the first improvement made over the model presented in \cite{webb12}. Recent $N$-body simulations by \citet{webb13} have shown that the historical assumption that tidal radii are imposed at perigalacticon is invalid because a GC is able to fill its instantaneous $r_t$ at all times, independent of its orbital eccentricity. More specifically, the mass normalized limiting radius of a cluster ($r_{L,n}=\frac{r_L}{M^{\frac{1}{3}}}$) is the same at a given $R_{gc}$, independent of cluster orbit.

However, comparing the instantaneous $r_t$ of a cluster to its observationally determined $r_L$ is also incorrect. \citet{kupper10} found that the bulk of the cluster, and hence the surface brightness profile, is nearly constant over an orbital period and more accurately reflects the mean tidal field that the cluster experiences. So while the true $r_L$ of a cluster changes with orbital phase \citep{webb13}, the observational limiting and effective radius as determined by a \citet{king62} model does not. To best compare with observations we need to calculate the effective radii of our simulated clusters, as the effective radius does not fluctuate as dramatically with orbital phase \citep{kupper10, webb13} and will therefore be more comparable to observationally determined effective radii.

\begin{figure} 
\centering
\includegraphics[width=\columnwidth]{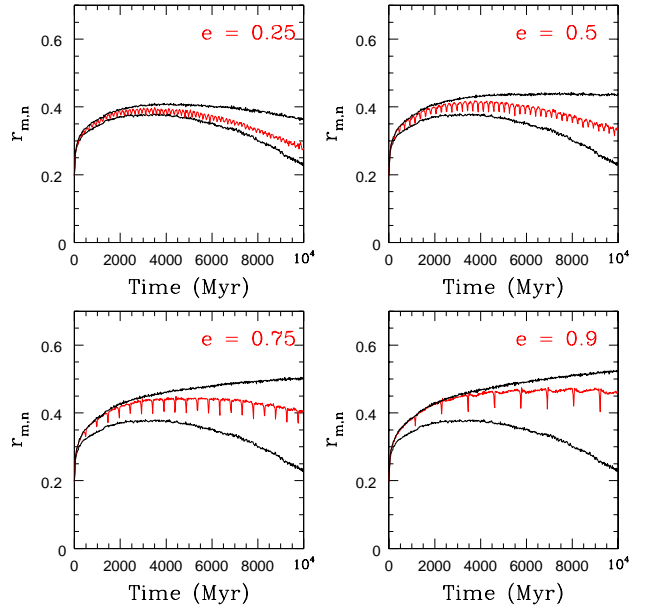}
\caption{Mass normalized $r_m$ of simulated star clusters on eccentric orbits (red) compared to clusters with circular orbits at $R_p$ (lower black line, always 6 kpc) and $R_{ap}$ (upper black line) as a function of time. Data taken from \citet{webb13}.}
  \label{fig:rm_nbody}
\end{figure}

In Figure \ref{fig:rm_nbody} we plot the mass normalized half-mass radii $r_{m,n}=\frac{r_m}{M^{\frac{1}{3}}}$ of various $N$-body model clusters as a function of time. A detailed discussion of the $N$-body models presented here is done in \citet{webb13}. With the infinite resolution of our model clusters, $r_h$ can fluctuate dramatically from time step to step. Therefore we use the half-mass radius $r_m$ to trace the evolution of $r_h$ as it remains consistent between time-steps. Even though $r_m$ is always slightly larger than $r_h$, the two radii scale the same with respect to time \citep{webb13}. In each panel, the lower black line is for a model cluster with a circular orbit at 6 kpc. The red line is for a model cluster with an eccentric orbit that has a perigalactic distance of 6kpc. Clusters were modelled with eccentricities of 0.25, 0.5, 0.75, and 0.9, with the eccentricity marked in each panel. The upper black line in each panel is for a model cluster with a circular orbit at the apogalactic distance of the eccentric cluster, which in these cases are 10 kpc, 18 kpc, 43 kpc, and 104 kpc. 

In Figure \ref{fig:rm_nbody}, the $r_{m,n}$ profile of clusters with circular orbits (black lines) decreases smoothly over time. Clusters with eccentric orbits (red lines) only undergo a brief fluctuation at $R_p$, but are also more or less smooth from one time-step to the next. A smooth evolution in $r_m$ is in agreement with the results of \citet{kupper10} discussed above. The effective radius also appears to be linked to the time-averaged tidal field that the cluster experiences: highly eccentric clusters are closer in size to clusters with orbits at $R_a$ while clusters with low eccentricities are comparable to clusters with circular orbits at $R_p$.

In order to predict $r_m$ or $r_h$ given the orbit and limiting radius of a cluster, we note that $r_{m,n}$ increases strongly as a function of eccentricity in Figure \ref{fig:rm_nbody}. Hence the $r_{m,n}$ of two clusters with the same $R_p$ and at the same $R_{gc}$ will not be the same if they have different orbits. From the results of our $N$-body simulations in \citet{webb13} (Figure \ref{fig:rm_nbody}) as well as larger mass versions of each model (presented in \citet{leigh13}), we find that the ratio of $r_{m,n}$ for a cluster with an eccentric orbit to $r_{m,n}$ for a cluster with a circular orbit at $R_p$ increases linearly with eccentricity after 10 Gyr. More specifically, clusters with eccentric orbits have effective radii that are a factor of $(1+0.31 \times e)$ larger than if they had circular orbits at $R_p$. The uncertainty in the correction factor of 0.31 is $\pm 0.01$. The correction factor is applicable to old GCs, but further simulations are required to determine how it depends on a GCs evolutionary stage.

In order to determine the effective radius of each simulated cluster, we first  calculate their tidal radii as if they had a circular orbit at $R_p$ given the formalism of \cite{bertin08}. The derivation of $r_t$ by \cite{bertin08} is ideal as it makes no assumptions regarding the potential of the host galaxy except that it must be spherically symmetric. Therefore the mass profile of M87 determined by \citet{mclaughlin99} can be used to determine the galactic potential. We next assume that all clusters are tidally filling, such that $r_L$ can be set equal to $r_t$ at perigalacticon. We explore the effects of non-tidally filling clusters in Section~\ref{s_underfill}. The perigalactic effective radius ($r_h$ assuming a circular orbit at $R_p$) is then calculated given the central concentration of the cluster and assuming that it can be represented by a \citet{king62} model. The true $r_h$ will be a factor of $(1.0+0.31\times e)$ larger than the perigalactic case.

\subsection{Including Orbital Anisotropy}

The second major improvement to our model involves  the anisotropy parameter $\beta$. In our previous work \citep{webb12}, $\sigma_\theta$ and $\sigma_\phi$ were kept equal to the observed line of sight velocity dispersion when $\beta < 0$, while $\sigma_R$ was decreased based on Equation \ref{beta}. Similarly for $\beta > 0$,  $\sigma_R$ was kept equal to the observed line of sight velocity dispersion while $\sigma_\theta$ and $\sigma_\phi$ were decreased. This approach did have the desired effect of altering the distribution of cluster orbits, but the resulting velocity dispersion was no longer equal to the observed one. The improved simulation we use here now adjusts $\sigma_R$, $\sigma_\theta$ and $\sigma_\phi$ simultaneously such that Equation \ref{beta} is satisfied and the overall mean velocity dispersion equals the observed line of sight velocity dispersion.

\subsection{The Effect of Tidally Under-filling Clusters}

Previously we have assumed that all simulated clusters are \textit{tidally filling}, as it allows for a straightforward calculation of $r_L$ and $r_h$ for each cluster. But not all observed GCs are expected to be tidally filling \citep{gieles10}. Therefore we added the filling parameter $R_F = \frac{r_L}{r_t}$ to the simulation, where $r_L$ is the limiting radius (essentially, the observed outer radius) and $r_t$ is the theoretically permitted tidal radius. The simulation allows for all GCs to be tidally under-filling by the same amount ($R_F = constant$) in order to explore the effect that tidally under-filling GCs have on the exponent $\alpha$. With the exception of Section~\ref{s_underfill} $R_F$ is always set equal to 1.0.

\subsection{Observational Constraints}

Finally, we introduce a minimum $r_h$ cut-off set equal to the smallest measurable value from the resolution limit of our observations. The simulation already includes a tidal dissolution time and dynamical friction infall time cutoff of 10 Gyr as described in \citet{webb12}.

\section{Comparing Theory and Observations \label{sfour}}

To match the observations, populations of 10000 clusters were simulated following the real spatial profile such that the total number of clusters within 10 kpc of M87 is the same as the observed dataset. The ratio of number of blue clusters to red clusters was set equal to $3:2$, in agreement with the profiles in \citet{harris09b}. The only difference between red and blue clusters in our simulation is which radial distribution profile in Table \ref{table:gcsim} is used to determine cluster position.

\subsection{The Isotropic Case}

\begin{figure} 
\centering
\includegraphics[width=\columnwidth]{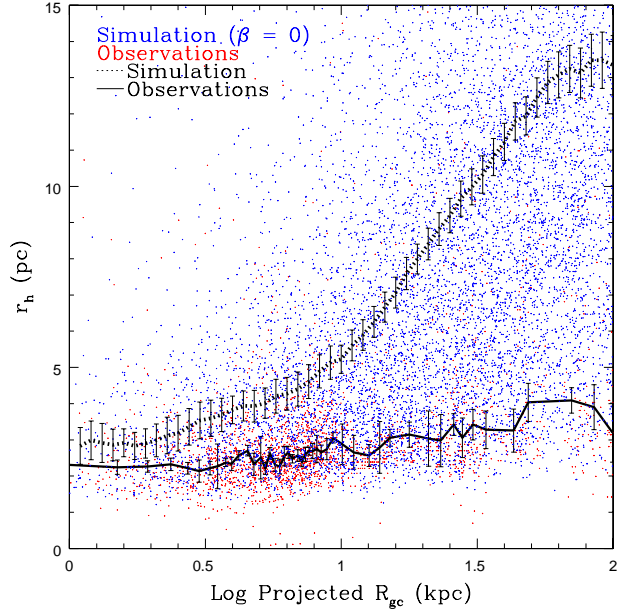}
\caption{$r_h$ and log $R_{gc}$ of each simulated GC (blue) for $\beta = 0$. The dashed black line marks the median $r_h$ calculated with radial bins containing 50 GCs each. For comparison purposes we also plot the observed clusters (red) and median (solid black line) from Figure \ref{fig:rh_obs}. Error bars represent the standard error $\sigma/\sqrt(n)$ as given by \citet{harris10}.}
  \label{fig:rh_B0compare}
\end{figure}

The first comparison between theory and observations was done for a model population with an isotropic distribution of orbits ($\beta=0$) and $R_F = 1$. The $r_h$ of both model (blue) and observed (red) clusters are plotted in Figure \ref{fig:rh_B0compare} as functions of $R_{gc}$.

Figure \ref{fig:rh_B0compare} indicates that an isotropic distribution of orbits produces a larger distribution of cluster sizes than observed, particularly at large $R_{gc}$. While the observations suggest $\alpha = 0.14$, the model predicts $\alpha = 0.41 \pm 0.01$, in closer agreement with basic tidal theory. Therefore this ''baseline" model strongly disagrees with the data, either in terms of the trend or the total scatter.

It should be noted that the assumption that all clusters are tidally filling is likely to be safest in the inner regions of the galaxy where the tidal field is strong \citep{alexander13}. Outer clusters, for which $r_t$ is considerably larger, are more likely to be tidally under-filling. The clear disagreement between the observations and the isotropic model suggest that either outer GCs are severely tidally under-filling, have preferentially radial orbits, or a combination of both.

\subsection{Anisotropic Cases} \label{s_anisotropy}

\begin{figure} 
\centering
\includegraphics[width=\columnwidth]{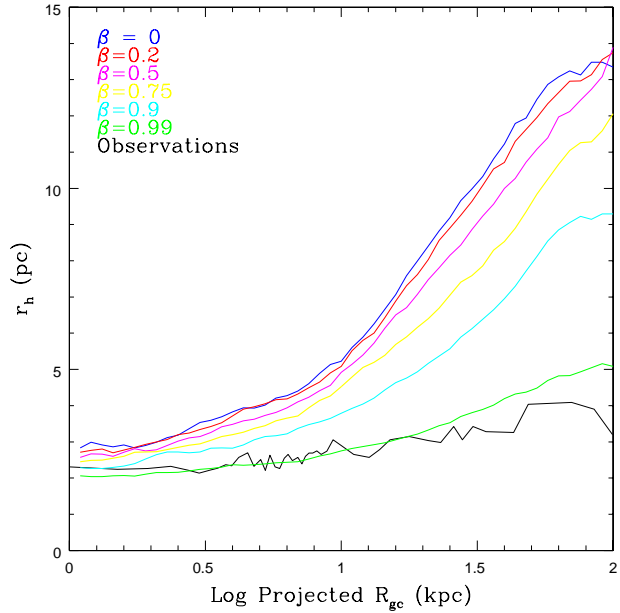}
\caption{Relationship between median $r_h$ and log $R_{gc}$ for simulated GC populations with different values of $\beta$. Median $r_h$ are calculated with radial bins containing 50 GCs each. The solid red line is the observed median effective radius From Figure \ref{fig:rh_obs}.}
  \label{fig:rh_Bcompare}
\end{figure}

\begin{figure} 
\centering
\includegraphics[width=\columnwidth]{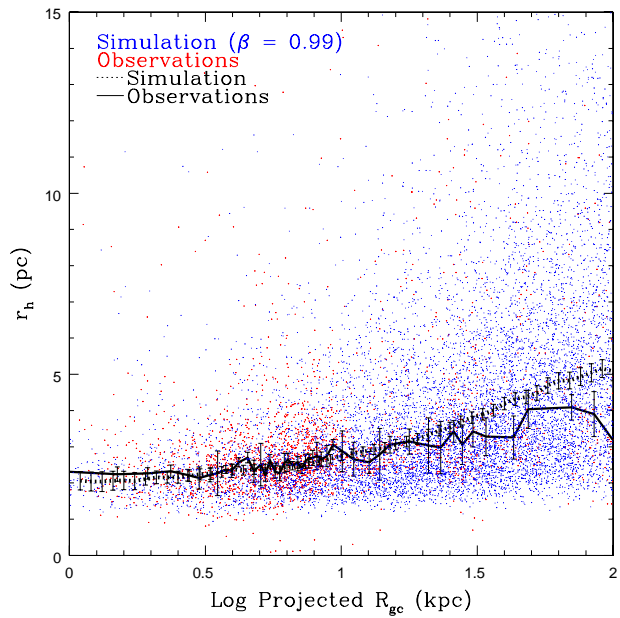}
\caption{$r_h$ and log $R_{gc}$ of each simulated GCs (blue) for a population with $\beta = 0.99$. The dashed black line marks the median $r_h$ calculated with radial bins containing 50 GCs each. For comparison purposes we also plot the observed clusters (red) and median (solid black line) from Figure \ref{fig:rh_obs}. Error bars represent the standard error $\sigma/\sqrt(n)$ as given by \citet{harris10}.}
  \label{fig:rh_B099compare}
\end{figure}

We first explore how much a non-isotropic distribution of orbits can minimize both the distribution of cluster sizes and the value of $\alpha$ in our model cluster population. In Figure \ref{fig:rh_Bcompare} we show the median $r_h$ as a function of galactocentric distance for models with different values of $\beta$. Very large values of $\beta$ are required in order to bring the median model cluster size down to the level of the observations. A $\beta$ of 0.99, which corresponds to a mean orbital eccentricity of 0.9, produces the closest agreement. In Figure \ref{fig:rh_B099compare}, which shows the actual distribution for $\beta=0.99$, the scatter in the simulated data points about the median line is greatly reduced and is more comparable to the observations than the $\beta = 0$ case. However the corresponding value of $\alpha$, equal to $0.21 \pm 0.01$, is still higher than the observed value of 0.14. Furthermore, while median cluster sizes are comparable in the mid to outer regions of M87, the $\beta = 0.99$ simulation underestimates cluster size in the inner regions of M87. These discrepancies suggest that $\beta$ likely increases with galactocentric distance. Previous observational and theoretical studies of M87 \citep{cote01, webb12}, NGC 3379 and NGC 821 \citep{weijmans09} , the Milky Way \citep{prieto08} and dark matter halos \citep{zait08, ludlow10} draw similar conclusions, although none of the existing data are consistent with such a high mean $\beta$.

Our simulation explicitly allows for the population to have an anisotropy profile $\beta(R_{gc})$. However, in order to put constraints on the profile as was done in \citet{webb12}, we first need to know the likely distribution of tidally filling and under-filling clusters in M87. Then the simulated $r_h$ profile will represent the observed profile as opposed to being an upper limit.

\subsection{The Effect of Tidally Under-filling Clusters}\label{s_underfill}

\begin{figure} 
\centering
\includegraphics[width=\columnwidth]{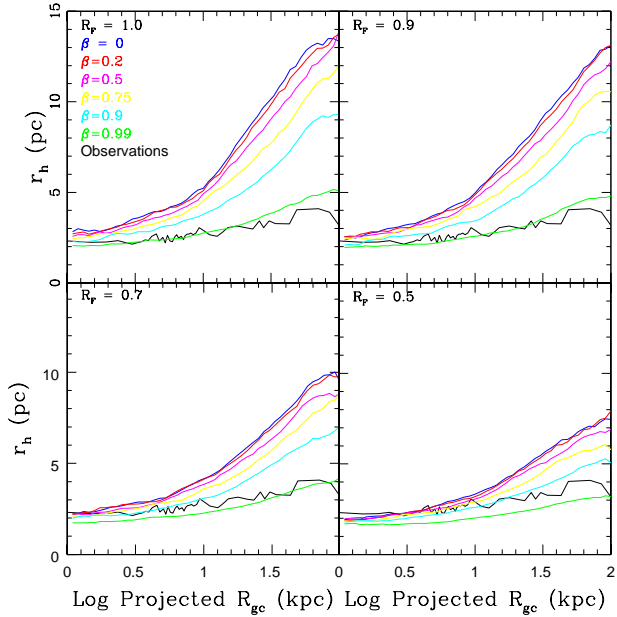}
\caption{Relationship between median $r_h$ and log $R_{gc}$ for simulated GC populations with different values of $\beta$ and $R_F = \frac{r_L}{r_t}$. Different values of $\beta$ are colour coded as indicated by the top left panel. The fraction by which clusters fill their tidal radii ($R_F$) is indicated in each panel. Median $r_h$ are calculated with radial bins containing 50 GCs each.}
  \label{fig:rh_rfill}
\end{figure}

We next explore how much the existence of tidally under-filling clusters can minimize both the distribution of cluster sizes and the value of $\alpha$ in our model cluster population. A recent study by \citet{alexander13} demonstrated that unless all clusters form tidally filling, a present day cluster population will be made up of a mix of tidally filling and under-filling clusters. They were able to reproduce a relationship between $r_h$ and $R_{gc}$ similar to the Galactic GCs by assuming the population formed under-filling and then evolved in a Milky Way-like potential. Allowing clusters to be tidally under-filling would not require such high values of $\beta$ as found in Section~\ref{s_anisotropy} or as steep an anisotropy profile. We illustrate this statement in Figure \ref{fig:rh_rfill} by simulating cluster populations with the same static values of $\beta$ as Figure \ref{fig:rh_Bcompare}, but with the filling parameter $R_F$ equal to $1$ (top left panel, same as Figure \ref{fig:rh_Bcompare}) , $0.9$ (top right panel), $0.7$ (bottom left panel), and $0.5$ (bottom right panel). While assuming all clusters under-fill their $r_t$ by the same amount must be unrealistic, it serves to illustrate the effect that under-filling clusters have on the relationship between $r_h$ and $R_{gc}$.

As clusters become more and more under-filling the median $r_h$ decreases at all galactocentric distances; similar to the effect of increasing $\beta$. Additionally, decreasing $R_F$ can also decrease the theoretical value of $\alpha$. Therefore some degeneracy exists between $\beta$ and $R_F$.

Realistically it is likely that clusters have a distribution in $R_F$, and that the distribution changes with $R_{gc}$. \citet{alexander13} found that the majority of inner clusters are tidally filling, while outer clusters range between tidally filling, near tidally filling, and tidally under-filling. The radial trend of clusters becoming tidally under-filling with $R_{gc}$ is also in agreement with observations of Galactic GCs. \citet{baumgardt10} found that inner GCs ($R_{gc} < 8$ kpc) were primarily tidally filling with $0.1 < \frac{r_h}{r_t} < 0.3$ while outer GCs ($R_{gc} > 8$ kpc) can be separated into two groups of tidally filling and tidally under-filling ($\frac{r_h}{r_t} < 0.05$) clusters. We will expand on this interpretation in a following paper.

\subsection{Red and Blue Globular Clusters}

Finally, we use our simulation to search for any evidence suggesting that the red and blue GCs in M87 may differ by more than just their radial distributions and metallicities. Observational works show that blue GCs have effective radii that are on average $20 \%$ ($\sim 0.4$ pc) larger than red GCs \citep[e.g.][]{kundu98, kundu99, larsen01, jordan05, harris09a, harris10, paolillo11, blom12, strader12, woodley12, usher13}. The size difference is also observed in our study, with mean blue cluster size being $28 \%$ ($\sim 1.0 $ pc) larger than the mean red cluster size. We suggest that the size difference we find is bigger than in other galaxies because our sample extends out to beyond 100 kpc: since clusters can reach large sizes in the outer regions of galaxy, and since the outer regions are dominated by blue clusters, the mean size difference will be larger due to the abundance of large blue clusters. Leading explanations of why this size difference exists suggest that red and blue clusters have different formation, dynamical and stellar evolution histories \citep[e.g.][]{kundu98, jordan04, jordan05, harris09a, sippel12, schulman12}. Here we explore the possibility that the size difference may be due to different orbital anisotropy profiles. 

\begin{figure} 
\centering
\includegraphics[width=\columnwidth]{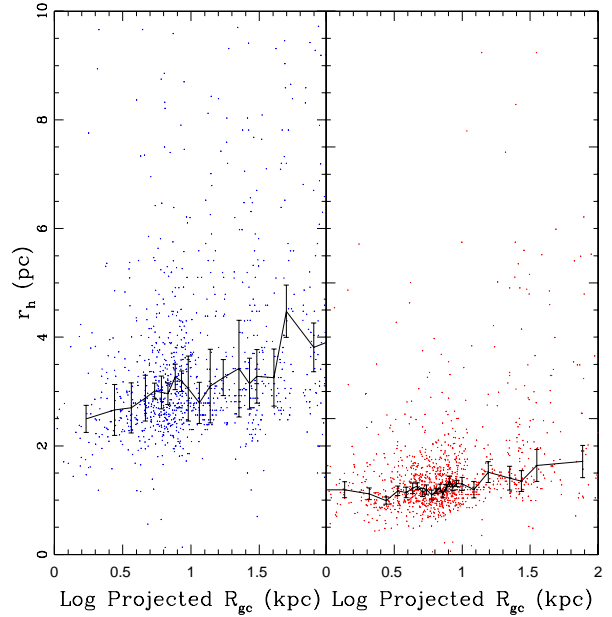}
\caption{$r_h$ versus log $R_{gc}$ for observed blue GCs (left panel) and red GCs (right panel). The solid black lines indicate the median $r_h$ for red and blue clusters respectively, and are calculated with radial bins containing 50 GCs each. Error bars represent the standard error $\sigma/\sqrt(n)$ as given by \citet{harris10}.}
  \label{fig:rh_rvsb}
\end{figure}

Figure \ref{fig:rh_rvsb} shows the sizes of the observed blue and red GC populations in the left and right panels. The blue and red populations have the same values of $\alpha$, equal to $0.11 \pm 0.01$ and $0.11 \pm 0.02$ respectively, but their $r_h$ profiles are offset by approximately 1 pc. The size difference does not change with $R_{gc}$, in agreement with recent studies \citep[e.g.]{usher13}. The different radial profiles of the red and blue clusters cause the global $\alpha \sim 0.14$ to be larger than the $\alpha$'s of the two sub-populations. 

The identical values of $\alpha$ but different mean $r_h$ between the red and blue populations cannot be explained by orbital anisotropy alone. The offset could be explained if outer red clusters are preferentially under-filling and have less eccentric orbits than outer blue clusters. If the blue population has been accreted from in-falling satellite galaxies then they should now be on highly eccentric orbits. Accreted blue clusters may also have larger $r_h$ than red clusters if the mean tidal field they experienced as a member of the satellite galaxy is weaker than the mean field experienced by red clusters. Our future study which combines the effect of orbital anisotropy and tidally under-filling clusters will shed more light on this issue.

\section{Summary and Conclusions \label{sfive}}

We present brand new HST observations of the halo regions of M87, and perform size measurements and photometry on all identified GCs. Combining this dataset with Archive images of the central regions of M87 allow us to probe the relationship between $r_h$ and $R_{gc}$ out to $R_{gc} \sim 100$ kpc with over 2000 GCs. We find that $r_h$ scales as $R_{gc}^{0.14}$, consistent with studies of most other giant E galaxies. We attempt to explain this very shallow relationship by invoking the presence of both orbital anisotropy and clusters that are tidally under-filling. To develop this interpretation we simulate many GC populations orbiting in the tidal field of M87, having a different orbital anisotropy parameter ($\beta$) or filling their tidal radii by different amounts.

Comparisons between our simulations and observations suggest that if all clusters are tidally-filling, inner clusters may have a near-isotropic distribution of orbits but outer clusters must have extremely radial orbits $\beta = 0.99$. Such high values of $\beta$ are not supported in the literature. 

However, allowing for the existence of tidally under-filling clusters relaxes the constraints on $\beta$ as tidally under-filling clusters serve both to decrease mean cluster size and flatten the theoretical relationship between $r_h$ and $R_{gc}$. We also apply these results to the red and blue cluster sub-populations separately to explain why blue clusters are on average larger than red clusters. In our observational dataset, red and blue clusters both scale as $r_h \propto R_{gc}^{0.11}$, but blue clusters are on average 1 pc larger. The only way we could theoretically reproduced this trend in our simulation is to assume outer red clusters are preferentially under-filling and have a more isotropic distribution of orbits.

Therefore, if both orbital anisotropy and the effect of tidally under-filling clusters are present in our simulation, we can reproduce the power-law proportionality between $r_h$ and $R_{gc}$ for both the cluster population as a whole and the red and blue cluster sub-populations. Future studies will employ the use of MCMC formalism to properly explore the degeneracy between increasing orbital anisotropy and tidally under-filling clusters, as both serve to decrease $r_h$. Furthermore, as previously indicated neither $\beta$ or $R_F$ are expected to be fixed values but are more likely functions of $R_{gc}$. 

The question of why orbital anisotropy is present in the cluster population and why some clusters are tidally under-filling remain open. Issues regarding whether or not initial cluster populations are under-filling and what portion of the present day population could have been accreted make constraints on the orbital anisotropy profile and filling parameter difficult. Nevertheless, all conclude that the evolution of clusters with different initial sizes and orbits as well as the accretion of satellite galaxies and their cluster populations are key to understanding the characteristics of present day cluster populations.

\section{Acknowledgements}

We would like to thank the referee for constructive comments and suggestions regarding the presentation of the paper. JW acknowledges support from the Dawes Memorial Fellowship for Graduate Studies in Physics. AS and WEH acknowledge financial support through research grants from the Natural Sciences and Engineering Research Council of Canada.


\end{document}